\documentclass[prb, twocolumn, showpacs, superscriptaddress, floatfix]{revtex4}
\usepackage{graphicx, subfigure}
\usepackage{dcolumn}

\begin{document}

\title{Classical potential describes martensitic phase transformations
  between the $\alpha$, $\beta$ and $\omega$ titanium phases.}

\author{R. G. Hennig}
\affiliation{Department of Material Science and Engineering, Cornell
  University, Ithaca, New York, 14853}
\affiliation{Department of Physics, Ohio State University, Columbus,
  Ohio 43210}
\author{T. J. Lenosky}
\affiliation{C8 Medisensors, San Jose, California 95124}
\author{D. R. Trinkle}
\affiliation{Department of Physics, Ohio State University, Columbus,
  Ohio 43210}
\affiliation{Department of Material Science and Engineering, University of
Illinois at Urbana-Champaign, Urbana, Illinois 61801}
\author{S. P. Rudin}
\affiliation{Los Alamos National Laboratory, Los Alamos, New Mexico, 87545}
\author{J. W. Wilkins}
\affiliation{Department of Physics, Ohio State University, Columbus,
  Ohio 43210}

\date{\today}

\begin{abstract}
  A description of the martensitic transformations between the
  $\alpha$, $\beta$ and $\omega$ phases of titanium that includes
  nucleation and growth requires an accurate classical potential.
  Optimization of the parameters of a modified embedded atom potential
  to a database of density-functional calculations yields an accurate
  and transferable potential as verified by comparison to experimental
  and density functional data for phonons, surface and stacking fault
  energies and energy barriers for homogeneous martensitic
  transformations.  Molecular dynamics simulations map out the
  pressure-temperature phase diagram of titanium.  For this potential
  the martensitic phase transformation between $\alpha$ and $\beta$
  appears at ambient pressure and 1200~K, between $\alpha$ and
  $\omega$ at ambient conditions, between $\beta$ and $\omega$ at
  1200~K and pressures above 8~GPa, and the triple point occurs at
  8GPa and 1200~K.  Molecular dynamics explorations of the dynamics of
  the martensitic $\alpha-\omega$ transformation show a fast-moving
  interface with a low interfacial energy of 30~meV/\AA$^2$.  The
  potential is applicable to the study of defects and phase
  transformations of Ti.
\end{abstract}


\maketitle

\section{Introduction}

Martensitic phase transitions control systems ranging from shape
memory alloys\cite{Otsuka98} to steels\cite{Martensite} to planetary
cores.\cite{Vocadlo03} They are diffusionless structural
transformations proceeding near the speed of sound.\cite{Martensite}
Martensitic transformations frequently appear in alloy design as a way
to improve materials properties, but their occurrence can also limit
materials performance.

Titanium's great technological importance\cite{Williams03} makes it an
ideal example for the development of physics-based predictive methods
for materials problems.  Titanium displays several phases as a
function of pressure and temperature.  At ambient conditions titanium
stabilizes in the hexagonal close-packed (hcp) $\alpha$ phase.  At
ambient pressure and temperatures above 1155~K the $\alpha$ phase
transforms to the high-temperature body-centered cubic (bcc) $\beta$
phase.  Under pressure the $\alpha$ phase transforms into the
hexagonal $\omega$ phase.\cite{Jamieson63} The high-pressure $\omega$
phase consist of a three atom hexagonal structure equivalent to
AlB$_2$ with Ti occupying sites of alternating triangular and
honeycomb layers.  The crystal structure of Ti at 0~K has not been
determined experimentally.  Extrapolation of the $\alpha$-$\omega$
phase boundary\cite{Sikka82} in Ti indicates $\omega$ as the ground
state phase.  Free energy calculations of $\alpha$ and $\omega$ within
the quasiharmonic approximation using a TB model show phonon entropy
stabilizing the $\alpha$ phase at ambient temperature.\cite{Rudin04}
Ti displays martensitic transformations between its $\alpha$ (hcp),
$\beta$ (bcc) and $\omega$ (hexagonal) phases.  Two-phase
$\alpha$/$\beta$ alloys make up many industrial titanium alloys such
as Ti-6Al-4V because the presence of $\beta$ phase in the $\alpha$
matrix improves strength and creep resistance.\cite{Williams03}
Titanium transforms from $\alpha$ to brittle $\omega$ under pressure
creating serious technological problems for $\beta$ stabilized
titanium alloys.  Impurities greatly affect the $\alpha$ to $\omega$
transformation; for example, as little as 1~at.\% oxygen in commercial
Ti alloys suppresses it.\cite{Vohra77,Gray92,Hennig05}

To understand these transformations in Ti and its alloys we begin with
the study of the phase transformations in pure titanium.  Our approach
to a theoretical understanding of these transformation involves three
steps: (1) find the homogeneous atomistic pathway of the martensitic
transformation in pure titanium, (2) use this pathway to estimate the
effect of impurities, and (3) determine the heterogeneous nucleation
and dynamics of the martensitic phase transformation.

The homogeneous transformation pathway for all three martensitic
transformations in Ti is known.  Burgers described the $\alpha$ to
$\beta$ transformation in Zr;\cite{Burgers34} the same mechanism
occurs in Ti.  The $\beta$ to $\omega$ transformation occurs via plane
collapse along the $[111]$ direction corresponding to the longitudinal
$\frac{2}{3} [111]$ phonon.\cite{Hatt60, Fontaine70, Persson00} More
recently Trinkle {\it et al.} determined the homogeneous pathway of
the $\alpha$ to $\omega$ transformation.\cite{Trinkle03, Trinkle05} A
systematic approach generated all possible pathways that were then
successively pruned by energy estimates using elastic theory,
tight-binding (TB) methods and density-functional theory (DFT).

The speed of the diffusionless martensitic transformation traps dilute
impurities, providing candidate pathways for alloyed materials.
Hennig {\it et al.} determined the effect of interstitial and
substitutional impurities on the $\alpha$ to $\omega$
transformation.\cite{Hennig05} DFT nudged-elastic band refinements
yield the change in both the relative stability of and the energy
barrier between the phases due to impurities.  The resulting
microscopic picture explains the suppression of the $\alpha$ to
$\omega$ transformation in commercial Ti alloys.

The final step involves studying the full atomistic dynamics of the
nucleation and growth of the martensitic phase; this requires
molecular dynamics simulations of large systems.  For the required
system sizes an accurate quantum mechanical treatment by DFT or TB
methods becomes too computationally demanding.  Such simulations call
for a classical potential to allow the treatment of appropriate length
and time scales for nucleation and growth of martensites.

In this paper we develop a classical potential of the modified
embedded atom method\cite{Lenosky00} (MEAM) type for Ti.  The
potential accurately describes the stability of the $\alpha$, $\beta$
and $\omega$ phases and is applied to study the dynamics of the
martensitic $\alpha$ to $\omega$ transformation and the interfacial
energy between the $\alpha$ and $\omega$ phase.
Section~\ref{Sec:Optimization} describes the calculations for the DFT
database, the functional form of the MEAM potential and the
optimization of the potential parameters to the DFT database.  The
accuracy of the potential is tested by comparing phonon spectra,
surface and stacking fault energies as well as energy barriers for
homogeneous martensitic transformations to DFT, TB and experimental
results.  In Section~\ref{Sec:Applications} we apply the potential to
study the phase diagram of Ti and the martensitic phase
transformations between the phases.  We estimate the interfacial
energy between $\alpha$ and $\omega$ and show that the classical MEAM
potential accurately describes the stability range of the three Ti
phases and the phase transformations between them.

\section{Optimization of the classical potential to density functional
 database}
\label{Sec:Optimization}

To describe the interactions between the Ti atoms and to enable
large-scale molecular dynamics simulations we develop a classical
potential.  The modified embedded atom method provides the form of the
potential\cite{Lenosky00} with potential parameters optimized to a
database of DFT calculations.  A second DFT database provides testing
data for the potential.  The optimization of the model proceeds
iteratively.  Systematically adding DFT results to the fitting and
testing databases improves the accuracy and extends the applicability
of the model.  This enables the development of a potential that
accurately reproduces the properties of all three Ti phases relevant
for the description of the martensitic phase transitions.  Available
experimental data confirms the accuracy of the resulting Ti MEAM
potential.

\subsection{Density functional calculations}

The DFT calculations are performed with {\sc
  Vasp},\cite{Kresse93,Kresse96b} a density functional code using a
plane-wave basis and ultrasoft Vanderbilt type
pseudopotentials.\cite{Vanderbilt90,Kresse94} The generalized gradient
approximation (GGA) of Perdew and Wang is used.\cite{PW91} A
plane-wave kinetic energy cutoff of 400~eV ensures energy convergence
to 0.3~meV/atom.  The $k$-point meshes for the different structures
are chosen to guarantee an energy accuracy of 1~meV/atom.  We treat
the Ti $3p$ states as valence states in addition to the usual $4s$ and
$3d$ states to provide an accurate treatment of the interaction at
close interatomic distances.

The DFT database for the fitting of the potential parameters consists
of energies, defects, forces and elastic constants for a variety of Ti
phases as well as energies of configurations along the TAO-1
transformation pathway from $\alpha$ to $\omega$.\cite{Trinkle03,
  Trinkle05} Relaxations determine the ground state energies and
lattice parameters of the $\alpha$, $\beta$, $\omega$, fcc, A15 and
simple hexagonal phases.  The volume dependence of the energy for
$\alpha$, $\beta$, $\omega$, fcc and A15 and the elastic constants of
the $\alpha$, $\beta$ and $\omega$ phases at their equilibrium volumes
are calculated.  Snapshots of short DFT molecular dynamics simulations
for $\alpha$, $\beta$ and $\omega$ at 800~K at their respective ground
state volumes with and without a vacancy defect provide force-matching
data.\cite{Ercolessi94} Defect structures and formation energies are
determined by relaxations for single vacancies and single interstitial
atoms in a 96-atom ($4\times 4\times 3$) supercell for $\alpha$ and a
108-atom ($3\times 3\times 4$) supercell for $\omega$ with a $2\times
2\times2$ $k$-point sampling grid.  This results in a 1~at.\% defect
concentration.  The atom positions are relaxed until the atomic-level
forces are smaller than 20~meV/\AA.

The DFT database for the test of the accuracy of the potential
consists of additional interstitial defects in $\alpha$ and $\omega$,
phonon spectra for $\alpha$, $\beta$ and $\omega$, surface energies
for $\alpha$ and $\omega$, and the I$_2$ stacking fault in $\alpha$.
The phonon calculation employ the direct force method and supercells
of 150 atoms ($5\times5\times3$) for $\alpha$, 125 atoms
($5\times5\times5$) for $\beta$, and 135 atoms ($3\times3\times5$) for
$\omega$. The surfaces are constructed by separating the crystal along
a high-index plane.  The surface energies result from relaxing a
periodic stacking of an 18~\AA\ to 20~\AA\ thick slab of rectangular
hcp cells with a 10~\AA\ vacuum region and a perfect bulk cell with
the same cell vectors.  A $k$-point mesh equivalent to $13 \times 13
\times 1$ is used for both the bulk and the slab calculation.
Relaxations of a $1\times1\times5$ supercell of $\alpha$ with and
without a single I$_2$ stacking fault determine the stacking fault
energy.

Comparison to available experimental data for phonon spectra, surface
energies, the stacking fault in $\alpha$, and the $p$-$T$ phase
diagram of Ti further confirm the accuracy of the potential as do
approximate energy barriers for homogeneous martensitic
transformations in TB.

\subsection{Modified embedded atom potential}

\begin{figure*}[t]
  \includegraphics[width=18cm]{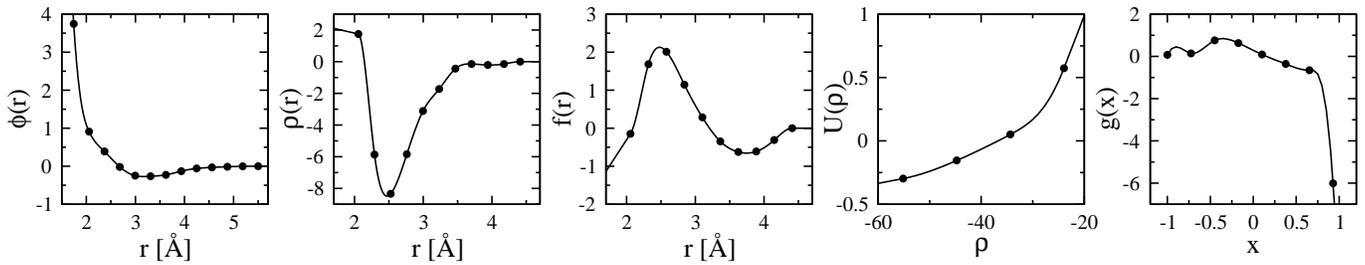}
  \caption{
    The five cubic splines of the MEAM potential.  The lines denote
    the spline interpolation between the spline knots represented by
    the solid circles.  The splines are linearly extrapolated beyond
    the last spline knot.}
  \label{fig:Splines}
\end{figure*}

The MEAM formalism was originally developed by Baskes\cite{Baskes87}
as an extension of the embedded-atom method.  The original MEAM
includes an angular-dependent electron density to model the effects of
bond bending; a series of four terms with $s$, $p$, $d$ and $f$
character describe the angular densities.  The original MEAM potential
has been applied to a variety of systems ranging from the
semiconductors Si\cite{Baskes87, Swadener02} and Ge\cite{Baskes89} to
bcc and fcc metals\cite{Lee00,Lee03} to several binary alloy
systems\cite{Baskes89, Baskes92} and recently to hcp Ti and
Zr.\cite{Kim06} While the Ti and Zr potentials accurately reproduce
properties of the hcp phase, they do not describe the bcc and $\omega$
phases.\cite{Kim06} Here we aim to develop a potential that accurately
describes all three phases and the transformations between them.

\begin{table}[t]
  \caption{Parameters specifying the five cubic splines that comprise the
    MEAM potential. The first part of the table lists the number of
    knots $N$ for each spline and the range of the spline variables
    $t_\mathrm{min}$ and $t_\mathrm{max}$.  
    The middle part of the table gives the values at equally
    spaced spline knots defined by $t_i=t_\mathrm{min} +
    i(t_\mathrm{max}-t_\mathrm{min})/N$ where N is the number of spline knots.
    Finally, the derivatives of the splines at their endpoints are listed
    in the last part of the table.}
  \label{tab:Splines}
  \begin{ruledtabular}
  \begin{tabular}[c]{l r r r r r}
    & \multicolumn{1}{c}{t} & \multicolumn{1}{r}{$t_\mathrm{min}$} & \multicolumn{1}{r}{$t_\mathrm{max}$} & N \\
    \colrule
    $\phi$ & \multicolumn{1}{c}{$r$[\AA]}            &  1.7427 &  5.5000 & 13 \\
    $\rho$ & \multicolumn{1}{c}{$r$[\AA]}            &  2.0558 &  4.4100 & 11 \\
    $f$    & \multicolumn{1}{c}{$r$[\AA]}            &  2.0558 &  4.4100 & 10 \\
    $U$    & \multicolumn{1}{c}{$\rho_\mathrm{tot}$} &-55.1423 &-23.9383 &  4 \\
    $g$    & \multicolumn{1}{c}{$\cos(\theta)$}      & -1.0000 &  0.9284 &  8 \\[0.5em]
    \colrule
    $i$ & \multicolumn{1}{r}{$\phi(r_i)$ [eV]} &
          \multicolumn{1}{r}{$\rho(r_i)$} &
          \multicolumn{1}{r}{$f(r_i)$} &
          \multicolumn{1}{r}{$U(\rho_i)$ [eV]} &
          \multicolumn{1}{r}{$g(x_i)$} \\
    \colrule
     0 &  3.7443 &  1.7475 & -0.1485 & -0.29746 &  0.0765 \\
     1 &  0.9108 & -5.8678 &  1.6845 & -0.15449 &  0.1416 \\
     2 &  0.3881 & -8.3376 &  2.0113 &  0.05099 &  0.7579 \\
     3 & -0.0188 & -5.8399 &  1.1444 &  0.57343 &  0.6301 \\
     4 & -0.2481 & -3.1141 &  0.2862 &          &  0.0905 \\
     5 & -0.2645 & -1.7257 & -0.3459 &          & -0.3574 \\
     6 & -0.2272 & -0.4429 & -0.6258 &          & -0.6529 \\
     7 & -0.1293 & -0.1467 & -0.6120 &          & -6.0091 \\
     8 & -0.0597 & -0.2096 & -0.3112 \\
     9 & -0.0311 & -0.1442 &  0.0000 \\
    10 & -0.0139 &  0.0000 \\
    11 & -0.0032 \\
    12 &  0.0000 \\[0.5em]
    \colrule
    $i$  & \multicolumn{1}{r}{$\phi'(r)$} &
         \multicolumn{1}{r}{$\rho'(r)$} &
         \multicolumn{1}{r}{$f'(r)$} &
         \multicolumn{1}{r}{$U'(\rho)$} &
         \multicolumn{1}{r}{$g'(x)$} \\
         & \multicolumn{1}{r}{[eV/\AA]} &
         \multicolumn{1}{r}{[\AA$^{-1}$]} &
         \multicolumn{1}{r}{[\AA$^{-1}$]} &
         \multicolumn{1}{r}{[eV]} &
         \multicolumn{1}{r}{}  \\
    \colrule
    0  & -20.0   & -1.0 & 2.7733 & 0.0078 &   8.3364 \\
    $N$&  0.0    &  0.0 & 0.0000 & 0.1052 & -60.4025 \\
   \end{tabular}
   \end{ruledtabular}
\end{table}

More recently, Lenosky {\it et al.} modified the original MEAM
potential by using cubic splines for the functional
form.\cite{Lenosky00} This removes the constraint of fixed angular
character and allows for additional flexibility of the potential.
Furthermore, the use of splines reduces the cost of evaluation over
the original functional form providing increased computational
efficiency.  In practice, the evaluation of the spline-based MEAM
potentials is only about a factor of two slower than that of EAM
potentials.  The spline-based MEAM was successfully applied to study
Si.\cite{Lenosky00, Birner01, Ciobanu04} This success of the
spline-based MEAM, its improved flexibility and its higher
computational efficiency motivate our use of this functional form
here.  The MEAM potential is implemented into two freely-available
large-scale parallel molecular dynamics codes.\cite{Goedecker02,
  ohmms}

The MEAM potential used in this work separates the energy into
  two parts\cite{Lenosky00}
\begin{equation}
  \label{eq:MEAM1}
  E = \sum_{ij} \phi(r_{ij}) + \sum_i U(\rho_i)
\end{equation}
with the density at atom $i$
\begin{equation}
  \label{eq:MEAM2}
  \rho_i =  \sum_j \rho(r_{ij}) + \sum_{jk} f(r_{ij}) f(r_{ik}) g(\cos(\theta_{jik})),
\end{equation}
where $\theta_{jik}$ is the angle between atoms $j$, $i$ and $k$
centered on atom $i$.  The functional form contains as special cases
the Stillinger-Weber\cite{Stillinger85} ($U(x) = x$ and $\rho = 0$)
and the embedded-atom (EAM) ($f = 0$ or $g = 0$) potentials.  The five
functions $\Phi(r)$, $U(\rho)$, $\rho(r)$, $f(r)$ and
$g(\cos(\theta))$ are represented by cubic splines.\cite{NumRec} This
allows for the necessary flexibility to accurately describe a complex
system such as Ti and provides the computational efficiency required
for large scale molecular dynamics simulations.

\subsection{MEAM Potential fit}

The spline parameters are optimized by a novel algorithm that involves
an extensive parameter search.  A detailed description of the
algorithm will be published separately.\cite{Lenosky07}
Figure~\ref{fig:Splines} and Table~\ref{tab:Splines} show the splines
and spline parameters of the best MEAM potential.

\begin{figure}[t]
  \includegraphics[width=8.5cm]{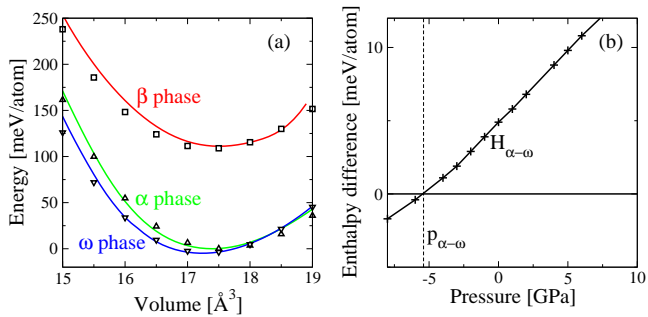}
  \caption{(color online) Fitted energies of $\alpha$, $\beta$ and
    $\omega$ Ti as a function of volume.  (a) The MEAM potential is
    fit to the energy of $\alpha$, $\beta$ and $\omega$ at several
    volumes.  The DFT results appear as symbols and the fitted MEAM
    curves as lines.  (b) The enthalpy difference between $\alpha$ and
    $\omega$ as a function of pressure for MEAM.  The MEAM potential
    predicts a transition pressure $p_{\alpha-\omega}$ of about
    $-5$~GPa at zero temperature similar to DFT which predicts the
    $\omega$ phase to be the ground state.\cite{Rudin04}}
  \label{fig:Volume}
\end{figure}

\begin{table}[t]
  \caption{Fitted energies relative to the $\alpha$ phase and
    lattice parameters for different Ti phases within the MEAM
    potential and DFT calculations (in parentheses).
    The cohesive energy of the $\alpha$ phase is 5.171~eV/atom in GGA
    and 4.831~eV/atom in MEAM.}
  \label{tab:Cohesive}
  \begin{ruledtabular}
  \begin{tabular}[c]
    {l@{\extracolsep{1em}}
     d@{\extracolsep{-1em}}d@{\extracolsep{-1em}}
     d@{\extracolsep{0em}}d@{\extracolsep{0.0em}}
     d@{\extracolsep{-1em}}d}
    Phase &
    \multicolumn{2}{l}{\hspace{-1.2em}$E_\mathrm{c}$[meV/atom]} &
    \multicolumn{2}{c}{$V$ [\AA$^3$]} &
    \multicolumn{2}{c}{$c/a$} \\
    \colrule
    $\alpha$  &   0 &   (0) & 17.40 & (17.55) & 1.596 & (1.583) \\
    $\omega$  &  -5 &  (-5) & 17.24 & (17.28) & 0.611 & (0.619) \\
    $\beta$   & 111 & (108) & 17.51 & (17.34) &       &         \\
    \colrule
    fcc       &  39 &  (58) & 17.83 & (17.53) &       &         \\
    A15       &  54 & (192) & 17.99 & (17.49) &       &         \\
    hexagonal & 396 & (353) & 17.25 & (17.78) & 0.982 & (0.999) \\
   \end{tabular}
   \end{ruledtabular}
\end{table}

Predictions of the resulting potential confirm its accuracy and
transferability.  Table~\ref{tab:Cohesive} compares the DFT energies
and lattice parameters with the MEAM values for the experimentally
observed $\alpha$, $\beta$ and $\omega$ phases and the fcc, A15 and
simple hexagonal structures.  The fitted MEAM values of the cohesive
energy of the $\alpha$ phase of 4.831~eV and the lattice parameter of
2.931~\AA\ agree closely with the experimental values (4.844~eV,
2.951~\AA) and the DFT results (5.171~eV, 2.948~\AA).  For the two
low-energy structures $\alpha$ and $\omega$ the final MEAM potential
is also fitted to the energy as a function of volume.  For $\beta$ and
fcc the fit includes the equilibrium lattice constants and energies
relative to hcp.  For the simple hexagonal structure the potential is
fitted to the energy relative to hcp for the structure with both
lattice parameters fixed to the DFT values.

\begin{table}[t]
  \caption{Fitted elastic constants in units of GPa for $\alpha$,
    $\beta$ and $\omega$ Ti for the MEAM potential compared
    to the DFT fitting data and experiment.  The MEAM potential
    accurately reproduces the DFT elastic constants.
    The experimental values for $\alpha$ Ti are measured at 4~K
    (Ref.~\onlinecite{Simmons71}) and the $\beta$ Ti elastic constants
    at 1238~K (Ref.~\onlinecite{Petry91}).  The deviation between
    the calculated and measured elastic constants for $\beta$ Ti
    stems from the high temperature needed to stabilize the structure.
    For $\alpha$ and $\omega$ Ti, internal relaxations are neglected in the
    MEAM and DFT calculations.}
  \label{tab:Elastic}
  \begin{ruledtabular}
  \begin{tabular}[c]{l c c c c c}
    & c$_{11}$ & c$_{12}$ & c$_{44}$ & c$_{33}$ & c$_{13}$ \\
    \colrule\\[-0.7em]
    \multicolumn{6}{c}{--- \emph{$\alpha$-phase} ---} \\[0.2em]
    MEAM & 174 &  95 & 58 & 188 & 72 \\
    GGA  & 172 &  82 & 45 & 190 & 75 \\
    Exp. & 176 &  87 & 51 & 191 & 68 \\[0.3em]
    \multicolumn{6}{c}{--- \emph{$\omega$-phase} ---} \\[0.2em]
    MEAM & 191 &  78 & 48 & 233 & 64 \\
    GGA  & 194 &  81 & 54 & 245 & 54 \\[0.3em]
    \multicolumn{6}{c}{--- \emph{$\beta$-phase} ---} \\[0.2em]
    MEAM &  95 & 111 & 53 & -- & -- \\
    GGA  &  95 & 110 & 42 & -- & -- \\
    Exp. & 134 & 110 & 36 & -- & -- \\
   \end{tabular}
   \end{ruledtabular}
\end{table}

\begin{table}[t]
  \caption{Comparison of fitted point defect energies in
    the $\alpha$ and $\omega$ phases for the MEAM potential
    with DFT and three tight-binding
    potentials.\cite{Trinkle06, Rudin04, Mehl02}
    The defect energies are measured relative to the corresponding $\alpha$
    or $\omega$ phases at about 1~at.\% defect concentration.
    The RMS errors of defect energies relative
    to GGA are given for both phases.  The
    tetrahedral interstitial and the
    dumbbell-[0001] oriented along the c-axis are close in
    structure such that the tetrahedral defect in MEAM and GGA
    relaxes into the dumbbell structure.   For the tight-binding
    potential by Rudin {\it et al.}\cite{Rudin04} these two defects
    and the tetrahedral  interstitial in $\omega$ 
    collapse, an explanation and fix for this problem
    can be found in Ref.~\onlinecite{Trinkle06}.  In the NRL potential the
    tetrahedral defect is unstable and relaxes to the octahedral defect.}
  \begin{ruledtabular}
  \begin{tabular}[c]{l d d d d d}
    Site &
    \multicolumn{1}{c}{MEAM} &
    \multicolumn{1}{c}{GGA} &
    \multicolumn{1}{c}{Trinkle\cite{Trinkle06}} &
    \multicolumn{1}{c}{Rudin\cite{Rudin04}} &
    \multicolumn{1}{c}{NRL\cite{Mehl02}} \\
    \colrule\\[-0.7em]
    \multicolumn{6}{c}{--- \emph{$\alpha$-defects} ---} \\[0.2em]
    Vacancy        & 2.24 & 2.03 & 1.81 & 1.92 & 1.51 \\
    Divacancy-AB\footnote[1]{Not included in potential fitting.}
                   & 4.00 & 3.92 & 3.82 & 3.68 & 3.73 \\
    Octahedral     & 2.64 & 2.58 & 2.89 & 2.55 & 1.31 \\
    Tetrahedral    & \multicolumn{1}{r}{\it dumb.} &
    \multicolumn{1}{r}{\it dumb.} & 2.86 &
    \multicolumn{1}{r}{\it coll.} &
    \multicolumn{1}{r}{\it octa.} \\
    Dumbbell-[0001]$^a$
                   & 2.21 & 2.87 & 2.81 & \multicolumn{1}{r}{\it coll.} & 1.81 \\[0.3em]
    RMS deviation  & 0.24 & \multicolumn{1}{c}{--} & 0.14 & 0.16 & 0.76 \\[0.5em]
    \colrule \\[-0.7em]
    \multicolumn{6}{c}{--- \emph{$\omega$-defects} ---} \\[0.2em]
    Vacancy A      & 2.78 & 2.92 & 2.85 & 3.25 & 2.99 \\
    Vacancy B      & 0.82 & 1.57 & 1.34 & 1.90 & 1.01 \\
    Octahedral     & 3.79 & 3.76 & 4.11 & 3.67 & 3.20 \\
    Tetrahedral    & 2.93 & 3.50 & 3.58 & \multicolumn{1}{r}{\it coll.} & 2.86 \\
    Hexahedral$^a$ & 3.31 & 3.49 & 3.86 & 4.37 & 3.20 \\[0.3em]
    RMS deviation  & 0.42 & \multicolumn{1}{c}{--} & 0.25 & 0.5 & 0.42 \\
  \end{tabular}
  \end{ruledtabular}
  \label{tab:Defects}
\end{table}

Figure~\ref{fig:Volume}(a) shows the result of the energy fit at
different volumes for $\alpha$, $\beta$ and $\omega$.  The DFT and
MEAM energies agree to about 5~meV/atom with closer agreement near the
energy minimum and slightly worse agreement at high compression.  The
MEAM potential reproduces the $\omega$ phase as the ground state and
places the $\alpha$ phase slightly higher in energy by 5~meV/atom in
agreement with DFT calculations.\cite{Rudin04}
Figure~\ref{fig:Volume}(b) shows the resulting enthalpy difference
between $\alpha$ and $\omega$.  At zero temperature the $\alpha$ phase
transforms to $\omega$ at a pressure of $-5$~GPa in the MEAM
potential.

Table~\ref{tab:Elastic} compares the elastic constants in MEAM with
DFT results and experiment for the $\alpha$, $\beta$ and $\omega$
phases.  For both methods the elastic constants are calculated
neglecting internal relaxations in the $\alpha$ and $\omega$ phases.
Accurate elastic constants are important for the correct description
of the long-range strain fields around dislocations and other defect
structures as well as in martensitic transformations.  The low RMS
deviation between MEAM and DFT elastic constants of 13\% and maximum
deviation of 29\% demonstrates the quality of the fit and indicates
the accuracy of the potential for the effects of strain in Ti.

\begin{figure*}
  \includegraphics[width=17cm]{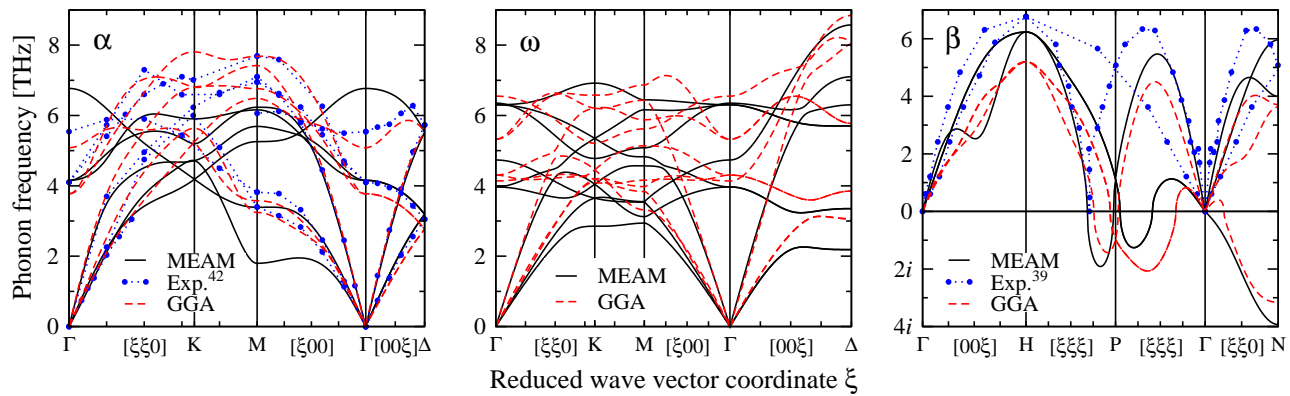}
  \caption{(color online) The phonon spectra of $\alpha$, $\beta$ and
    $\omega$ Ti for the classical MEAM potential (black solid lines)
    compared to DFT results (red dashed lines) and to experimental
    data for the $\alpha$ and $\beta$ phases (blue
    circles).\cite{Stassis79, Petry91} For the $\alpha$ phase the MEAM
    potential accurately reproduces the low energy acoustic branches
    reflecting the elastic constants.  For optical and high-energy
    acoustic branches the potential reproduces the experimental phonon
    frequencies within 25\%, generally underestimating the
    experimental values.  Similarly for the $\omega$ phase the
    acoustic branches of the MEAM potential closely match the DFT
    results with larger discrepancies of up to 20\% for the optical
    modes.  The $\beta$ phase is mechanically unstable\cite{Petry91}
    at low temperatures and transforms to $\alpha$ or $\omega$.  This
    instability is reflected in unstable (imaginary) phonon branches
    in the DFT and MEAM calculations.  The imaginary phonon for
    T-$[110]$ at the N-point corresponds to the $\beta$ to $\alpha$
    transformation mechanism.  The imaginary L-$\frac{2}{3}[111]$
    phonon is responsible for the $\beta$ to $\omega$ transformation.}
  \label{fig:Phonons}
\end{figure*}

Table~\ref{tab:Defects} compares the formation energies of point
defects in the $\alpha$ and $\omega$ phases for the MEAM potential
with DFT and several TB potentials.  The defect relaxations are
performed at fixed equilibrium volume for single defects in a $4\times
4\times 3$ and $3\times 3\times 4$ supercell of $\alpha$ and $\omega$,
respectively.  This corresponds to a defect concentration of 1\%.  The
formation energies of the various interstitial atoms and vacancies in
both phases agree well with the DFT results.  In fact, the RMS errors
of the energies are similar for the classical MEAM and the TB
potentials.  In addition the MEAM potential stabilizes the correct
defects found in DFT calculations.  The fitting data did not include
the $\alpha$ dumbbell-[0001], the $\alpha$ divacancy-AB, or the
$\omega$ hexahedral interstitial defects.  The close agreement for
these defects indicates the model's accuracy.

Molecular dynamics simulations for the defects confirm their stability
in MEAM.  Calculations for larger simulation cells with 1080 and 1296
atoms show that the residual finite-size error of the defect formation
energies is smaller than 0.1~eV.  Some interstitials can lower their
energy in MEAM by symmetry breaking.  In the $\alpha$ phase, the
octahedral interstitial moves to an off-center octahedral site with an
energy 0.5~eV lower than the central octahedral site. The dumbbell
cants, lowering its energy by 0.09~eV.  In $\omega$ the octahedral and
hexahedral interstitials reduce their energy by about 1 and 0.9~eV,
respectively, moving into an off-center position.  No attempt is made
to determine the defect stability in DFT or TB calculations.

Experimental values for defect formation energies in Ti are rather
difficult to obtain due to the presence of the $\alpha$-$\beta$
transformation and the sensitivity of diffusion to impurities such as
oxygen.  Based on an empirical relationship that connects the onset of
positron trapping with the vacancy formation energy, positron
annihilation experiments estimate for $\alpha$-Ti a value of $1.27\pm
0.05$~eV.\cite{Hashimoto84} Isotope diffusion measurements result in
diffusion activation energy of 1.75~eV.\cite{LandoltBornsteinIII26}
Assuming a vacancy mechanism of diffusion, the two measurements lead
to an estimate for the energy barrier of diffusion of about 0.5~eV.
Both experimental values are significantly smaller than DFT
predictions.  The origin of the discrepancy remains unclear and beyond
the scope of this paper.

The RMS force matching errors are 25\% ($\beta$), 27\% ($\alpha$) and
27\% ($\omega$).  The forces depend quadratically on the phonon
frequencies, hence, the expected error in the phonons is approximately
half the force-matching error and should be of the order of 15\% for
all three phases.  Errors in the low-frequency acoustic branches are
significantly less reflecting the accuracy of the elastic constants.

\subsection{Accuracy of the MEAM potential}
\label{Sec:Accuracy}

We test the accuracy of the MEAM potential by comparing the phonon
spectra, surface and stacking fault energies and energy barriers for
homogeneous martensitic transformations to DFT, TB and experiment.

\subsubsection{Phonons}

Figure~\ref{fig:Phonons} compares the phonon spectra obtained with the
MEAM potential for the three Ti phases $\alpha$, $\beta$ and $\omega$
with the available experimental data and results from DFT
calculations.  For all three phases, the MEAM potential reproduces the
GGA phonons within about 15\%, good agreement that can be attributed
to the force-matching method and fitting to elastic constants.  The
acoustic branches are better reproduced than the the optical modes,
reflecting the accuracy of the elastic constants.  Both acoustic and
optical phonons are needed to describe the shuffle and strain degree
of freedom in the martensitic phase transformations.

For the $\alpha$ phase the MEAM potential reproduces the overall trend
of the experimental phonon branches.  The optical L-[001] phonon at
$\Gamma$ is 20\% too high in MEAM and shows the wrong curvature away
from $\Gamma$.  For the K and M point, the MEAM potential
underestimates the experimental phonon frequencies by about 20 to
30\%.

No experimental data is available for the high-pressure $\omega$
phase.  We compare the phonon spectrum for the MEAM potential with
results from DFT calculations and find agreement between the MEAM
potential and the DFT results comparable to that of the $\alpha$ phase
with closely matching acoustic modes and larger deviations for the
optical branches.  In contrast to the $\alpha$ phase, for the $\omega$
phase the deviations for the optical branches are smaller and more
uniform across the Brilloin zone.

The high-temperature $\beta$ phase becomes mechanically unstable at
lower temperatures and shows a soft mode in the experimental data for
the L-$\frac{2}{3}[111]$ phonon.  The zero-temperature phonon results
for the MEAM potential and DFT reflect this instability in an unstable
(imaginary) phonon branch.  This mode is responsible for the $(111)$
plane collapse mechanism of the $\beta$ to $\omega$ transformation.
In addition, MEAM and DFT show an unstable phonon branch in the
T-$[110]$ direction at the N-point which corresponds to the Burgers
mechanism of the $\beta$ to $\alpha$ transformation.

\subsubsection{Surface and stacking fault energies}

Large changes of coordination number provide a challenging test for
atomistic potentials.  Specifically relevant to experiment are tests
on free surfaces.  Relaxations of low-index surfaces of the $\alpha$
and $\omega$ phases with MEAM and DFT determine the accuracy of the
classical potential here: Calculations for increasingly larger slabs
show that a slab thickness of more than 15~\AA\ results in surface
energies accurate to about 1~meV/\AA$^2$.

Table~\ref{tab:Surface} compares the surface energies of the $\alpha$
and $\omega$ phases in MEAM with DFT calculations.  The overall
agreement of the MEAM surface energies with the DFT values for both
phases is quite remarkable considering the fact that free surfaces
were not used to optimize the potential.  The average MEAM surface
energy is about 20\% too small.  The good agreement encourages the
potential's application to model systems with free surfaces such as
voids or cracks.

\begin{table}[t]
  \caption{Surface energies of the $\alpha$ and $\omega$ phase in
    MEAM compared to DFT calculations.}
  \label{tab:Surface}
  \begin{ruledtabular}
  \begin{tabular}[c]{l c c c c c c}
    & \multicolumn{3}{c}{$\alpha$}
    & \multicolumn{3}{c}{$\omega$} \\
    $\sigma$ (meV/\AA$^2$) & $[11\bar 20]$ & $[1\bar1 00]$ & $[0001]$ 
    & $[11\bar 20]$ & $[1\bar 1 0 0]$ & $[0001]$ \\
    \colrule
    MEAM       & 105 &  97 &  92 &  98 & 115 & 116 \\
    GGA        & 117 & 153 & 121 & 152 & 136 & 133 \\
   \end{tabular}
   \end{ruledtabular}
\end{table}

Stacking faults in hcp materials alter the structural sequence of
atomic planes in the c-direction.  Their energies test the accuracy of
the potential under changes of bond direction and second nearest
neighbor coordination.  There are three basic stacking faults possible
in hcp materials.  Intrinsic stacking faults $I_1$ and $I_2$ change
the hcp stacking sequence $ABAB$ to $ABAB|CBCB$ and $ABAB|CACA$,
respectively.  Extrinsic stacking faults introduce additional layers
in the hcp stacking sequence such as $ABAB|C|ABAB$.  The intrinsic
stacking fault $I_2$ describes a crystal sheared by a partial lattice
vector while both the intrinsic stacking fault I$_1$ and the extrinsic
stacking fault require a diffusive process.  In hcp materials the
$I_2$ stacking fault energy determines the dissociation of dislocation
on the basal plane into partial dislocations\cite{Girsheck98,
  Zaefferer03} and has been measured for $\alpha$-Ti.\cite{Patridge67}

The MEAM potential correctly predicts a metastable $I_2$ stacking
fault with a high stacking fault energy (170~mJ/m$^2$), though not as
high as in DFT (320~mJ/m$^2$) and experiment
(300~mJ/m$^2$).\cite{Patridge67} The high stacking fault energy in
MEAM would result in a narrow splitting of dislocations.  Elasticity
predicts a splitting on the order of 7~\AA\ (two Burgers vectors) for
a basal dislocation; this is consistent with a prediction of prismatic
slip, as expected for Ti.\cite{Bacon67, Bacon02} Since the MEAM
potential predicts a small dislocation splitting and reproduces the
elastic constants, the potential may provide an accurate description
of dislocation interactions at short and long distances.

\subsubsection{Energy barrier from $\alpha$ to $\omega$}

\begin{figure}
  \includegraphics[width=8.0cm]{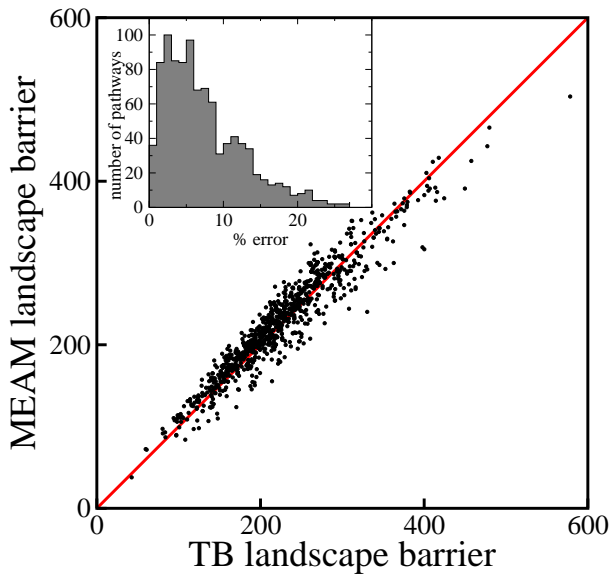}
  \caption{(color online) Comparison of energy barriers of different
    pathways for the $\alpha$ to $\omega$
    transformation~\cite{Trinkle03, Trinkle05} with MEAM and
    TB.\cite{Trinkle06} The inset shows the distribution of the
    deviations between the TB and the MEAM potential.  The RMS
    deviation is 5.5\% and the maximum deviation 27\% over the set of
    977 pathways.}
  \label{fig:Barrier}
\end{figure}

Figure~\ref{fig:Barrier} compares the energy barriers for different
mechanism of the $\alpha$ to $\omega$ transformation in MEAM and the
TB potential of Trinkle {\it et al.}\cite{Trinkle06} The energy
barrier is calculated for a two-dimensional reduced phase space for
all the pathways considered in Refs.~\onlinecite{Trinkle03} and
\onlinecite{Trinkle05}.  The first degree of freedom describes the
strain of the $\alpha$ cell into the $\omega$ cell and the second
degree of freedom describes the shuffle motion of all atoms from their
$\alpha$ to $\omega$ position within the cell.  The energy barriers
predicted by MEAM agree within 27\% with the highly accurate TB
potential by Trinkle {\it et al.}\cite{Trinkle06} for all pathways and
show an RMS deviation of only 5.5\%.  This excellent agreement
indicates a highly accurate representation of the energy landscapes by
the MEAM potential.

\section{Application to phases and phase transformations}
\label{Sec:Applications}

The demonstrated high accuracy of the MEAM potential and its
computational efficiency enable medium to large scale predictive
simulations for titanium.  This section presents molecular dynamics
simulations for the different Ti phases and for an interface between
the $\alpha$ and $\omega$ phases.  The simulations determine the phase
stability and pressure-temperature phase diagram of $\alpha$, $\beta$
and $\omega$ Ti and the interfacial energy and mobility for the
$\alpha$--$\omega$ martensitic transformation.

\subsection{Equilibrium phase diagram of titanium}

Molecular dynamics simulations determine the stability of the Ti
phases as a function of pressure and temperature.  The simulations use
the TPN (constant temperature, constant pressure, constant number)
ensemble\cite{Parrinello82} with a time step of 1~fs.  To estimate the
stability range of the $\alpha$, $\beta$ and $\omega$ phases we
perform molecular dynamics starting from a cubic cell of $\beta$ with
432 atoms, that is comensurate with all three phases if properly
strained.  For each pressure and temperature value we simulate up to
1~ns and observe the phase evolution of the system.  Simulations for a
solid-liquid interface containing 864 atoms estimate the melting
temperature of the $\beta$ phase.

\begin{figure}
  \includegraphics[width=8.5cm]{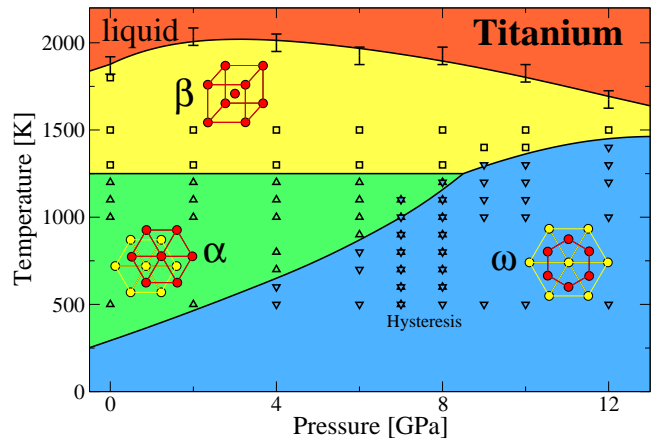}
  \caption{(color online) Equilibrium phase diagram of Ti as a
    function of pressure and temperature.  The MEAM potential
    accurately describes all three solid phases, $\alpha$, $\beta$ and
    $\omega$, of Ti and captures the martensitic phase transformations
    between them.  Near the triple point, the $\alpha$, $\beta$ and
    $\omega$ phases are nearly degenerate.  As a result, whether the
    $\beta$ phase transforms to either $\alpha$ or $\omega$ is
    controlled by the initial conditions of the molecular dynamics
    simulations.}
  \label{fig:Phasediagram}
\end{figure}

Figure~\ref{fig:Phasediagram} shows the predicted Ti equilibrium phase
diagram as a function of pressure and temperature for the classical
MEAM potential.  At pressures below 7~GPa and temperatures below
1200~K the $\beta$ phase transforms into the $\alpha$ phase by a shear
and shuffle motion of the atoms.  At pressures above 8~GPa and
temperatures below about 1300~K the $\beta$ phase transforms into the
$\omega$ phase.  The transition temperature between the $\alpha$ and
$\beta$ phase is nearly independent of the pressure while the
transition temperature for the $\beta$ to $\omega$ transition
increases with pressure.  The triple point between the $\alpha$,
$\beta$ and $\omega$ phase occurs at about 8~GPa and 1200~K. At zero
pressure the $\beta$ phase melts near 1900~K.  The melting temperature
first increases with pressure up to about 2000~K at 4~GPa and then
slowly decreases with pressure.

\begin{figure*}
  \includegraphics[width=16.2cm]{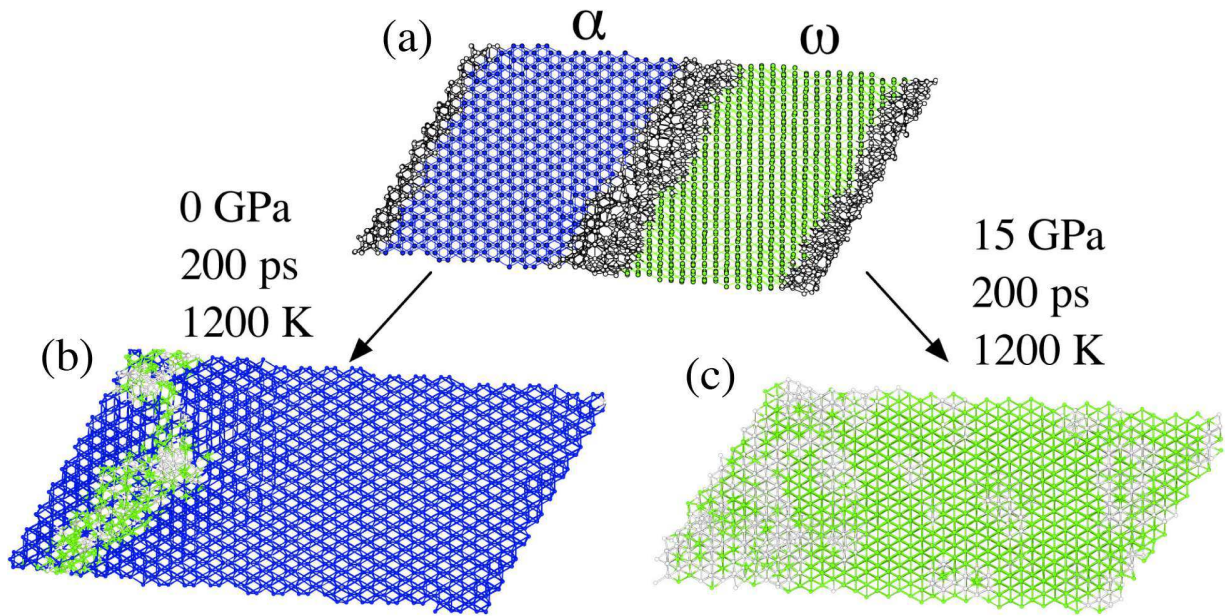}
  \caption{(color online) Molecular dynamics simulations of the
    $\alpha$-$\omega$ martensitic transformations.  (a) The relaxed
    $\alpha$-$\omega$ interface is created using the TAO-1 pathway and
    relaxed to produce the initial interfaces.  The energy for the
    interface is 30~meV/\AA$^2$.  Unidentified atoms are colored gray,
    while $\alpha$ and $\omega$ atoms are colored blue and green,
    respectively.  (b) Transformed geometry at zero compression.
    Molecular dynamics simulations of the relaxed interface structure
    at 1200~K for 200~ps transform the entire $\omega$ phase region.
    When the structure is fully relaxed to zero temperature, it is
    identified as entirely $\alpha$, except for the small region where
    the two interfaces merge and leave behind a number of defects.
    (c) Transformed geometry at 10\% volume compression.  The relaxed
    interface structure is first compressed by 10\% (corresponding to
    $\sim$10~GPa) and then simulated at 1200~K for 200~ps.  The entire
    $\alpha$ region transforms to $\omega$, except for a number of
    point defects (interstitials and vacancies).  For illustration,
    the slab is rotated around the horizontal axis so that the $c$
    axis of $\omega$ is perpendicular to the page.}
  \label{fig:hcp_omega}
\end{figure*}

The phase diagram of the MEAM potential agrees closely with
experimental observations.  The $\alpha$-$\beta$ transition occurs in
experiment at 1155~K compared to 1250~K in the simulation.  The
$\beta$ phase melts at 1943K in close agreement with the MEAM value of
1900~K.  Measurements of the $\alpha$-$\omega$ transformation pressure
show a large hysteresis with a transformation onset ranging from 2.9
to 9.0~GPa.\cite{Bundy63, Zilbershtein73, Zilbershtein75, Vohra77} The
accepted equilibrium transformation pressure of $2.0\pm0.3$~GPa was
estimated from samples under shear stress that reduce the
hysteresis.\cite{Zilbershtein75} Experimental values for the triple
point range from 8~GPa to 9~GPa and 900~K to 1100~K,\cite{Bandi66,
  Kutsar75} similar to the MEAM potential values of 8~GPa and 1200~K.
The close agreement of the MEAM phase diagram with experimental data
enables quantitative simulations for the martensitic phase
transformations between the $\alpha$, $\beta$ and $\omega$ phases.

\subsection{Pressure-induced martensitic phase transformations}

Under pressure Ti transforms from $\alpha$ to $\omega$ via the TAO-1
mechanism.\cite{Trinkle03, Trinkle05} As a first step towards a
detailed understanding of the nucleation, we simulate
$\alpha$-$\omega$ interfaces under compression at finite temperature.
The growth of a nucleus of a daughter phase in a parent phase is
controlled by the mobility of the interface at finite temperature
under a driving force.  For the Ti $\alpha$ to $\omega$ martensitic
phase transformation the free enthalpy difference between the two
phases provides the driving force.

We construct $\alpha$-$\omega$ interfaces using the TAO-1 supercell to
study the dynamics of the martensitic phase transformation.  We set up
interfaces between periodic slabs of untransformed TAO-1 supercells
($\alpha$ phase) and transformed TAO-1 supercells ($\omega$ phase)
that minimize lattice mismatch and strain while retaining periodicity.
The resulting interfaces are consistent with the pathway, minimize
mismatch at the boundaries and minimize strain in each phase.  The
simulation cell has periodic boundary conditions in all three
dimensions and contains a total of 3,600 atoms, half in each of the
two phases.  The system consists of alternating $\alpha$ and $\omega$
layers of 100~\AA\ thickness.  Relaxations of the initial interface
estimate an interfacial energy between $\alpha$ and $\omega$ of
30~meV/\AA$^2$, roughly a third of the calculated surface energies
(see Tab.~\ref{tab:Surface}).

All simulations are performed with {\sc Ohmms}\cite{ohmms} using a
frozen cell geometry and a Langevin thermostat to produce a constant
temperature of 1200~K.  A time step of 1~fs is used for all numerical
integration with the velocity-Verlet propagator.

Figure~\ref{fig:hcp_omega} shows the results of the molecular dynamics
simulations for this cell at 1200~K.  The simulations are performed at
two different volumes: zero compression for the $\omega$$\to$$\alpha$
transformation corresponding to approximately 0~GPa and 15\%\ volume
compression for the $\alpha$$\to$$\omega$ transformation corresponding
to about 15~GPa.  At both pressures the interface between $\alpha$ and
$\omega$ is mobile.  At 0~GPa the system transforms completely to
$\alpha$ and at 15~GPa completely to $\omega$, both within only
200~ps.  In both cases the interfaces between $\alpha$ and $\omega$
approach each other and partially annihilate, leaving behind a number
of interstitial and vacancy defects.

Temperature alone does not drive the transformation.  We performed
runs using the $\alpha$ and the $\omega$ slabs by themselves at 1200~K
with no compression and 10\%\ compression for 1~ns.  In both cases the
initial structure remained for the duration of the simulation.  This
indicates that homogeneous nucleation is not likely on the time scale
of nanoseconds in cells with a few thousand atoms, while the motion of
the interface does occur on such short time scale in these cells.

\section{Conclusion}

We developed and tested a classical potential for the complex phase
transformations of the technologically important Ti system.  The
potential is of the modified embedded atom form ensuring computational
efficiency, with parameters optimized to density functional
calculations.  The optimized potential describes the structure and
energetics of all three phases of Ti, the $\alpha$, $\beta$ and
$\omega$ phases.  The elastic constants, phonon frequencies, surface
energies and defect formation energies closely match density
functional results even when these were not included in the fitting
procedure.

Molecular dynamics simulations of the phase stability determine the
potential's equilibrium phase diagram in close agreement with
experimental measurements.  Simulations for the mobility of an
$\alpha$--$\omega$ interface demonstrate a high interfacial mobility
corresponding to the martensitic character of the $\alpha$--$\omega$
transformation.  The potential enables quantitative studies of point
defect evolution, grain boundary structures and mobility, as well as
phase transformations in the Ti system.

\begin{acknowledgments}
  This research is supported by DOE Grant No. DE-FG02-99ER45795 and
  under Contract No.~W-7405-ENG-36.  Computational resources were
  provided by the Ohio Supercomputing Center, NCSA, NERSC and PNL.
\end{acknowledgments}


\end{document}